\documentclass[12pt]{iopart}
\usepackage{graphicx}\usepackage{epsfig}\usepackage{color}
\usepackage{subfigure}\usepackage{lineno}
\begin{document}
%
\title[Thermal properties of light nuclei from $^{12}$C+$^{12}$C
  ...]
{ Thermal properties of light nuclei from
$^{12}$C+$^{12}$C fusion-evaporation reactions}
%
\author{L Morelli$^{1}$, G Baiocco$^{1,2}$~\footnote{Present address:
Dipartimento di Fisica dell'Universit\`{a} and INFN, Pavia, Italy}, 
M D'Agostino$^{1}$,
F Gulminelli$^2$, \\ M Bruno$^{1}$, U Abbondanno$^3$, S Appannababu$^4$,
S Barlini$^{5,6}$, \\ M Bini$^{5,6}$, G Casini$^6$, M Cinausero$^4$, M Degerlier$^7$,\\
D Fabris$^8$, N Gelli$^6$, F Gramegna$^4$, V L Kravchuk$^{4,9}$,\\
T Marchi$^{4}$, A Olmi$^{6}$, G Pasquali$^{5,6}$, S Piantelli$^{6}$, 
S Valdr\'e$^{5,6}$ \\and Ad R Raduta$^{10}$ }
\address{$^1$Dipartimento di Fisica ed Astronomia dell'Universit\`{a} and INFN,
Bologna, Italy}
\address{$^2$LPC (IN2P3-CNRS/Ensicaen et Universit\'e),
F-14076 Caen c\'edex, France}
\address{$^3$INFN Trieste, Italy}
\address{$^4$INFN, Laboratori Nazionali di Legnaro, Italy}
\address{$^5$Dipartimento di Fisica ed Astronomia dell'Universit\`{a},
  Firenze, Italy}
\address{$^6$INFN Firenze, Italy}
\address{$^7$ University of Nevsehir, Science and Art Faculty, Physics
  Department, Nevsehir, Turkey}
\address{$^8$INFN, Padova, Italy}
\address{$^9$National Research Center \textquotedblleft Kurchatov
  Institute\textquotedblright, Moscow, Russia}
\address{$^{10}$NIPNE, Bucharest-Magurele, POB-MG6, Romania}
\ead{luca.morelli@bo.infn.it}
%
%
\begin{abstract}
The $^{12}$C+$^{12}$C reaction  at 95 MeV has been studied through the
complete charge identification of its products
by means of the GARFIELD+RCo experimental
set-up at INFN Laboratori Nazionali di Legnaro (LNL).
In this paper, the first of a series of two,
a comparison to a dedicated Hauser-Feshbach calculation
allows to select a set of dissipative
events which corresponds, to a large extent, to the statistical
evaporation of highly excited $^{24}$Mg. Information
on the isotopic distribution of the evaporation residues in coincidence
with their complete evaporation chain is also extracted.
The set of data puts strong constraints on the behaviour of the level
density of light nuclei above the threshold for particle emission. In
particular, a fast increase of the level density parameter with
excitation energy is supported by the data.
Residual deviations from a statistical behaviour are seen in two
specific channels, and tentatively associated with a contamination from
direct reactions and/or $\alpha$-clustering effects.
These channels are studied in further details in the second paper of
the series. 
\end{abstract}
\noindent{\it NUCLEAR REACTIONS  12C(12C,X), E = 95 AMeV, 
Measured Fusion-evaporation reactions, Observed deviation from
statistical behaviour.\/} 

\submitto{\JPG}

\maketitle

\section{Introduction}

The statistical theory of Compound Nucleus (CN) decay is one of the oldest
achievements of nuclear physics and has proved its remarkable
predictive power since sixty years~\cite{ghoshal}.
Within this theory the detailed output of a generic nuclear reaction
is uniquely predicted under the knowledge of
nuclear ground state properties and level densities. The knowledge of level
densities is not only important for the understanding of nuclear
structure~\cite{alhassid}, but it is also required for different
applications of nuclear physics, from nucleosynthesis calculations to
reactor science. Its direct measurement from transfer
reactions~\cite{oslo} is limited to a relatively low excitation
energy domain. Above the thresholds for particle decay, level
densities are only accessible in evaporation reactions through the
theory of CN decay.

Despite the interest of the issue, mainly inclusive experiments have
been used up to now to constrain this fundamental
quantity~\cite{charity}, and very few studies exist altogether
concerning the evaporation of very light nuclei in the mass region
$A\approx 20$~\cite{report,indians,egidy}.
However, this mass region is very interesting to explore. Indeed
some excited states of different nuclei in this mass region are known
to present pronounced cluster structures.
These correlations may persist in the ground state along some selected
isotopic chains~\cite{neff}, and according to the Ikeda
diagrams~\cite{ikeda} alpha-clustered excited states are massively
expected at high excitation energies close to the multi-alpha decay
threshold in all even-even $N=Z$ nuclei. These cluster structures
have been evidenced in constrained density functional
calculations~\cite{maruhn,ebran,schuck} close to the threshold energy
of breakup into constituent clusters and even beyond.
They should lead to exotic non-statistical decays with a privileged
break-up into the cluster constituents which start to be identified
in the recent literature~\cite{freer,cluster}.

Such effects might be experimentally seen as an excess of cluster production
with respect to the prediction of the statistical model, provided that the
ingredients of the latter are sufficiently constrained via experimental data.
It is important to recall that the final inclusive yields represent integrated
 contributions over the whole evaporation chain.
Because of that, the information they bear on specific excitation energy
regions of the different nuclei explored during the evaporation process may
be model dependent~\cite{wallner} unless the decay chain is fully controlled
in a coincidence experiment.
To progress on these issues, we have performed an exclusive and
complete detection of the different decay products
emitted in $^{12}$C+$^{12}$C dissipative reactions at $95$ MeV.
We compared the experimental data to the results of a dedicated
Hauser-Feshbach code for the evaporation of light systems (HF$\ell$
from now on) with transmission coefficients and level
densities optimized in the $A\approx 20$ region~\cite{baiocco,prc}.

In this paper, the first of a series of two,
we show that all the observables of dissipative events are fully
compatible with a standard statistical behaviour,
with the exception of $\alpha$-yields in coincidence with Carbon and
Oxygen residues.

The good reproduction of a large set of inclusive and exclusive observables
by the statistical model allows to constrain the least known part of the theory,
namely the behaviour of the level density at high excitation energy,
well above the threshold of particle decay. 
We will show that a fast increase of the level density parameter
in the $A\approx 20$ mass region from $a\approx 2.4$ MeV$^{-1}$ at the
neutron separation energy, to $a\approx 3.5$ MeV$^{-1}$ at
$E^*/A\approx 3$ MeV is compatible with our data. 

The observed residual anomalies are tentatively attributed to clustering
effects which appear to survive even in the most dissipative events.
These effects will be studied in greater detail in the second paper of
this series.

\section{The statistical decay code}
\label {The statistical decay code}

In this section we give the main features of the Monte Carlo HF$\ell$
statistical decay code. For further details, see~\cite{baiocco}.\\
The evaporation of light particles is treated with the standard
Hauser-Feshbach (HF) formalism of CN decay~\cite{hauser-feshbach},
with n, p, d, t, $^3$He, $\alpha$ particles and $^6$Li,
$^7$Li emission channels included.
The expression for the decay width in channel $\xi$ for a hot nucleus
$(A,Z)$ excited in its state $C$ (specified by the energy $E^*$ and
the angular momentum $J$), in the framework of the HF model reads:

\begin{eqnarray}
\Gamma^C_{\xi}=\frac{1}{2\pi\rho_C} \int_0^{E^*-Q} d\epsilon_{\xi}
\sum_{J_d}  \ \sum\limits_{j=\vert J-J_d\vert}^{J+J_d} \ \sum\limits_{\ell =\vert j-s_p\vert}^{j+s_p}
T_{j,s_p}^{J}(\epsilon_{\xi}) \rho_d
\label{hf}
\end{eqnarray}

where $\epsilon_{\xi}$ is the relative kinetic energy of the decay
products (the daughter nucleus, labeled by $d$, and the evaporated
particle, labeled by $p$); $Q$ is the decay $Q$-value; $J_d$, $s_p$ 
and $\ell$ are the angular momentum of the daughter nucleus, the spin
of the evaporated particle and the orbital angular momentum of the
decay, respectively and summations 
include all angular momentum couplings between the initial and final
states; $T$ is the transmission coefficient; $\rho_{C}(E^*,J)$ and
$\rho_{d}(E^*-Q-\epsilon_{\xi},J_d) $ are the nuclear
Level Density (LD) of the decaying and of the residual nucleus,
respectively.\\ 
The widths $\Gamma^C_{i}$ are calculated for all possible decay
channels and the Branching Ratio (BR) associated with a specific channel
$\xi$ is obtained as the ratio between $\Gamma^C_{\xi}$ and the total
decay width for the hot nucleus:
$BR^C(\xi)=\Gamma^C_{\xi}/\sum_{i}\Gamma^C_{i}$.
This decay probability constitutes the main ingredient of the Monte
Carlo simulation.

In the case of the very light CN studied in this work,  
simple analytical expressions can be safely employed for the
transmission coefficients.
In our code we have adopted the empirical work of~\cite{gelbke}:

\begin{equation}
\label{tdiscr}
T_{\ell}(\epsilon_{\xi})=\frac{1}{1+\exp{\left(\frac{V_{b}-\epsilon_{\xi}}{\delta\cdot V_{b}}\right)}}
\end{equation}

where the barrier $V_{b}$ is a sum of a Coulomb and a centrifugal
term depending on $\ell$, hence on all coupled angular momenta, see
(\ref{hf}). Its full expression reads:

\begin{eqnarray}
V_{b}&=&\frac{1.44}{r_Z} \ \frac{Z_p(Z-Z_p)}{(A-A_p)^{1/3}
+A_p^{1/3}}\nonumber + \frac{\hbar^2 \ell(\ell+1)}{2r_Z^2}\frac{\frac{A}{A_p(A-A_p)}}{\left[(A-A_p)^{1/3}+A_p^{1/3}\right]^2}
\end{eqnarray}

The two free parameters $\delta$ and $r_Z$ were optimized to reproduce the decay of
discrete resonances~\cite{gelbke}. They depend
on the charge $Z_p$ of the evaporated particle, and $\delta$ also
depends on whether the emission takes place in the sub- or
above-barrier region.

Concerning the kinematics of the decay with angular momentum, we have adopted
the semi-classical approach proposed by the GEMINI
code~\cite{gemini}. 
Angular momenta are considered
as classical vectors, and \textbf{$\bf{j}_{a}$} and
\textbf{$\bf{j}_{b}$} are coupled under the assumption of
equiprobability for the module of their sum $\bf{j}_s$ between $\vert
\bf{j}_a-{j}_b\vert \le {j}_s  \le \vert {j}_a+{j}_b\vert$. 
Once the decay channel has been selected, the angular momentum $J_d$
is obtained through a 
maximization of $\rho_d(J)$ as a function of $J$.
Decay $Q$-values are calculated from experimental binding energies taken from
the Audi and Wapstra compilation~\cite{wapstra}.
Finally, a special effort has been devoted to the implementation of
the level density model.
In particular, all information on measured excited levels from the
online archive NUDAT2~\cite{nudat} has been explicitly and coherently included in
the decay calculation.

\subsection{The level density model}

The back-shifted Fermi gas model (BSFG), with the level density
parameter and the pairing backshift left as free fit parameters, is
known to be a phenomenological approach well suited to reproduce
the many-body correlated nuclear level density:
pairing effects are included through the backshift $\Delta_p$, and
all correlations are taken into account in the renormalization of the
LD parameter $a(E^*)$. 

In~\cite{egidy} level density
parameters for the BSFG model have been determined for a large set of
nuclei (310 nuclei between $^{18}$F and $^{251}$Cf), by the fit of
complete level schemes at low excitation energy and $s$-wave neutron
resonance spacings at the neutron binding energy.

In~\cite{egidy}, the adopted expression for $\rho(E^*)$ (after
integration on  angular momentum $J$ and parity $\pi$) reads:
 \begin{equation}
 \rho(E^*)=\frac{\exp[2\sqrt{a(E^*-E_2)}]}{12\sqrt{2}\sigma a^{1/4}(E^*-E_2)^{5/4}}
 \label{rhou}
 \end{equation}
where $\sigma$ is the spin cut-off parameter:
 \begin{equation}
\sigma^2 =0.0146A^{5/3}\frac{1+\sqrt{1+4a(E^*-E_2)}}{2a}
 \end{equation}
The energy backshift $\Delta_p=E_2$ is left as the first free
parameter in the data fitting. The second fit parameter is the
asymptotic value $\tilde{a}$ of the following functional form for
$a(E^*,Z,N)$~\cite{ignatyuk}:
\begin{equation}
a=\tilde{a}\left[1+\frac{S(Z,N)-\delta E_p}{E^*-E_2}\left(1-e^{-0.06(E^*-E_2)}\right)\right]
\label{aU}
\end{equation}
where $S(Z,N)=M_{exp}(Z,N)-M_{LD}(Z,N)$ is a shell correction term,
$M_{exp}$ and $M_{LD}$ being respectively the experimental mass and
the mass calculated with a macroscopic liquid drop formula
for the binding energy not including any pairing or shell corrections.
$\delta E_p$ is a pairing term expressed in terms of the deuteron separation energy.
Full details on the parameter definition and fit procedure can be
found in~\cite{egidy}. As a final result, analytic formulas for 
$E_2$ and $\tilde{a}$ as a function of tabulated nuclear properties
are given.
\begin{figure}
\centering
\includegraphics[angle=0, width=.6\columnwidth]{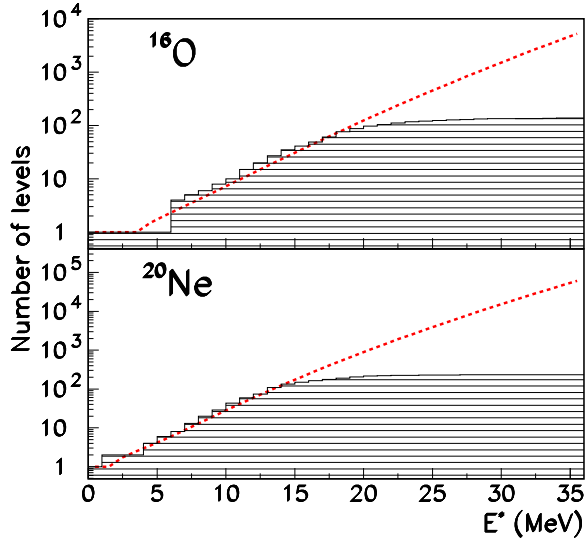}
\caption{Color online: Comparison between the cumulative number of
  levels given by (\ref{rhou}) (lines), and the cumulative counting
  of experimentally  measured levels from the NUDAT2 archive~\cite{nudat}
  (histograms).
}
\label{level_density}
\end{figure}
With such formulas for the calculation of LD
parameters, the model of (\ref{rhou}) allows for a very good
reproduction of experimental distributions of measured levels in the
mass region of interest for the present work.
Two selected examples are given in Figure~\ref{level_density}.
The isotope $^{20}Ne$ belongs to the fitted data set,
and the good agreement between the line and the histogram shows the
quality of the fit procedure of~\cite{egidy}.
Concerning $^{16}O$, the values of the parameters are an extrapolation of the
formulas proposed in~\cite{egidy} out of the fitted
data set; from the figure it is also clear that (\ref{rhou}) can be
considered reliable also for nuclei whose level density has not been
directly optimized. A similar agreement is observed for all the other
particle-stable isotopes in the mass region of interest for the present
study.\\
Still, numerical values for the pairing backshift and for the
asymptotic limit of $a(E^*)$ with increasing excitation energy
obtained through this approach are to be considered reliable only up
to $E^*/A\approx 1$ MeV for $A \approx 20$ nuclei. \\
In particular, it is found that the values of the level density
parameter needed to reproduce the information on discrete levels are
usually lower than the ones coming from higher-energy constraints,
through the reproduction of data for fusion-evaporation or
evaporation-after fragmentation studies ($E^*/A \approx 2 \div 3$ MeV).
A functional form giving a good reproduction of evaporation spectra at
very high excitation energy was proposed in~\cite{toke}:

\begin{equation}
a_{\infty}=\frac{A}{14.6}\left(1+\frac{3.114}{A^{1/3}}+\frac{5.626}{A^{2/3}}\right)
\label{aA}
\end{equation}

To correctly reproduce at the same time the low- and high-energy
experimental constraints, we have adopted a functional form for the
level density parameter that gives a continuous interpolation
between (\ref{aU}) and (\ref{aA}).

We have adopted the following expression:
\begin{equation}
 a(E^*,A)=\left\lbrace
\begin{array}{lll}
 a_D = (\ref{aU}) & & \ \ if \ \ E^*\le E_{m}+E_2\\
 a_C = \alpha \exp[-\beta (E^{*}-E_2)^2]+ a_{\infty}
 & & \ \ if \ \ E^* > E_{m}+E_2 \\
\end{array}
\right.
\label{am}
\end{equation}

The choice of a rapid (exponential) increase is imposed by the fact
that the asymptotic value (\ref{aA}) is 
connected to the opening of the break-up or multifragmentation
channels, which is a sharp threshold phenomenon.
The $\alpha$ and $\beta$ parameters are fully determined by the
matching conditions between the low-energy and high-energy regime:
$a_D(E_{m},A)=a_C(E_{m},A)$ and $a_C(E_l,A)=a_{\infty}\pm 10\%$.

Here, $E_l$ represents the limiting energy
at which the break-up or fragmentation regime is attained,
while $E_{m}$ is the excitation energy marking the transition
between the discrete and the continuum part of the spectrum.
This latter quantity  is of the order of
 $E_{m} \approx 10$ MeV, coherently with the value of the critical
energy for the damping of pairing effects in~\cite{abla}.
 In the case of light nuclei for which a large set of
measured levels is available, this value well corresponds to
the excitation energy maximizing the number of levels in bins of
$E^*$. Above $E_{m}$ the experimental information is too poor to
consider the set of resolved levels exhaustive of the nuclear level
density,  due to the physical emergence of the continuum.\\
\begin{figure}
\centering
\includegraphics[angle=0, width=.5\columnwidth]{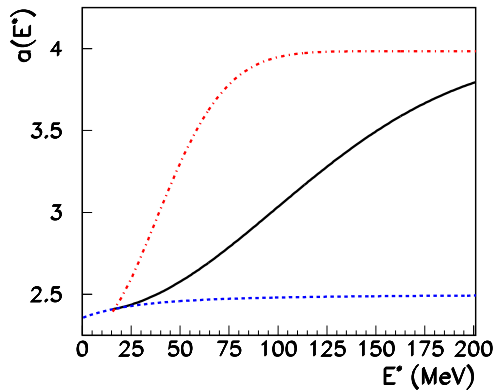}
\caption{(Colour online)  Level density parameter calculation for $^{20}Ne$.
Blue dashed line:  (\ref{aU}).
Black solid and red dot-dashed lines: (\ref{am}) with  $ E_l =$ 8 and 3 AMeV,
respectively.
}
\label{fig_elim}
\end{figure}
The limiting energy $E_l$  is then left as the only free parameter
of the calculation, governing the rapidity of the variation of
$a(E^*)$ above $E_{m}$. As an example of the overall functional
form resulting for the level density parameter, in Figure~\ref{fig_elim}
we plot $a(E^*,A)$ for $^{20}$Ne, for two different choices of
$E_l$.

In the statistical code, starting from a given CN
$(A,Z,J,E^*)$, the decay pattern is calculated with the Monte-Carlo
technique as a sequence of two-body decays governed by the emission
probability given by (\ref{hf}).
When the emitted particle leaves the daughter nucleus at an excitation
energy $E^*_d<E_m$, the excitation energy is considered as a discrete
variable, and one of the tabulated levels~\cite{nudat} of the daughter
discrete spectrum is populated.
The level is chosen according to the Breit-Wigner distribution of the
discrete levels considering their respective widths,
including the full spectroscopic information of~\cite{nudat}, and
the particle kinetic energy is adjusted if necessary
to ensure energy conservation.
When a particle bound level is populated, the subsequent decay is assumed to be due to a single $\gamma$
emission to the corresponding ground state. 
If the daughter excitation energy $E^*_d$ is greater than $E_m$, the
spectroscopic 
information is not sufficient to fully constrain the spin and energy
of the daughter nucleus. If measured excited states exist, they are
populated with a probability given by the ratio between the measured
level density from discrete states and the total level density
including the continuum states and given by (\ref{rhou}) with $a$
given by (\ref{am}). If no levels are known, the emission is
assumed to take place in the continuum.

\section{Experiment and data selection}
The measurement was performed in the third experimental Hall of LNL.
The $^{12}$C+$^{12}$C reaction had been already studied in a previous
experiment, and first results on the persistence of cluster correlations
in dissipative reactions highlighted by the comparison of the data set
with HF$\ell$ calculations were reported~\cite{prc}.
The limited statistics of the experiment prevented detailed studies of
the breakup angular and energy correlations. Here we report the
analysis of the new data-taking, which confirms our previous findings
and additionally allows to study the deviations from statistical
behaviour in specific channels and in greater detail.

A pulsed beam (less than 2 ns FWHM, 400 ns repetition period) of
$^{12}$C provided by the TANDEM accelerator impinged with a
self-supporting $^{12}$C target, with a thickness of $85\ \mu g/cm^2$.
The bombarding energy was 95 MeV.

\subsection{The experimental setup}
\label{The experimental setup}
The experimental setup is composed by the GARFIELD detector, covering
almost completely the angular range of polar angles from 30$^{\circ}$
to 150$^{\circ}$, and the Ring-Counter (RCo) annular detector~\cite{garf},
centered at $0^{\circ}$ with respect to the beam direction and
covering forward laboratory angles in the range
$5^{\circ}\leq \theta \leq 17^{\circ}$.

The combination of the two devices allows for a nearly-$4 \pi$  coverage
of the solid angle, which, combined with a
high granularity, permits to measure the charge, the energy and the
emission angles of nearly all charged reaction
products.
The setup also provides
information on the mass of the emitted charged products in a wide
range of particle energy~\cite{epj}.

The GARFIELD apparatus is a two-detection stage device, consisting of two
microstrip gaseous drift chambers ($\mu SGC$),
filled with $CF_4$ gas at low pressure (50 mbar) and placed back to
back, with $CsI(Tl)$ scintillation detectors
lodged in the same gas volume.

Due to the small size of the studied system, mainly light
  particles are emitted in the reaction which are efficiently detected
  and identified through the use of the $fast-slow$ shape
  method for the 180 $CsI(Tl)$ scintillators~\cite{fs}.

The energy identification thresholds result, on average, 3, 6, 9,
20, 7 MeV for $p$, $d$, $t$, $^3$He, and $\alpha$ particles,
respectively. As for other experimental devices using the $fast-slow$
technique~\cite{Rivet}, $^3$He can be discriminated from $\alpha$'s
starting from $\approx$~20 MeV.
This increase of the $^3$He threshold does not affect too much the
$\alpha$ yield in our reaction, since $^3$He is estimated to represent
less than 2-3\% of Z=2 particles~\cite{prc}.
In all the experimental percentages, the associated error takes
into account both the statistical error and the possible $^3He-\alpha$
contamination.
In the present analysis, the  information coming from the $\mu$SGC has
been used to validate the particle identification, especially in the
lower part of the range, where the $fast-slow$ curves tend to merge. 

The RCo detector is an array of three-stage telescopes realized in a
truncated cone shape.
The first stage is an ionization chamber (IC), the second a $300 \mu$m
reverse mounted $\rm Si(nTD)$ strip detector, 
and the last a $CsI(Tl)$ scintillator.

The angular resolution is $\Delta\theta\approx \pm 0.7^{\circ}$ and
the energy resolution of silicon strips and $CsI(Tl)$ detectors
resulted 0.3\% and 2-3\%, respectively. 
In the present experiment, reaction products with $Z
\ge 3$ have relatively low energies and
are stopped in the $Si$ detectors. Therefore, they can be identified
only in charge thanks to the $\Delta E-E$
correlation between the energy loss in the gas and the residual energy
in the silicon detectors, with 1 AMeV energy threshold.
Only for the high energy tails of $3 \le Z \le 5$ fragments mass
identification has been possible, thanks to  the application of a
pulse shape technique to signals coming
from the $Si$ detectors~\cite{faziaps}.
Light charged particles (LCP, $Z=1, 2$) flying at the RCo angles and punching
through the $300\ \mu m$ Si pads ($E/A \ge$ about  $6$ MeV) are
identified in charge and mass by the conventional Si - CsI $\Delta E - E$
method. LCP stopped in the silicon stage are identified only in
charge. 

More details on this setup can be found in~\cite{epj}.

\subsection{Minimum bias compound nucleus selection}
\label{Minimal bias event selection}
The analysis considers only events with a coincidence
between at least one LCP, detected and identified
in GARFIELD, and a particle or fragment ($Z \ge 3$)
detected at forward angles in the RCo and identified
in charge. In the case of a fusion-evaporation reaction, this latter
is the residue $Z_{res}$ of the CN decay chain, and it
is expected to have a velocity close to the center-of-mass velocity of
the reaction, $v_{CM}\approx 2$ cm/ns.
Due to the lack of isotopic resolution for such low energy fragments, a
hypothesis on their mass has to be done. Our initial hypothesis
(to be further discussed in \S\ref{Calorimetry and isotopic distributions})
is $A_{res}=2\cdot Z_{res}$.

\begin{figure}
\centering
\includegraphics[angle=0, width=.5\columnwidth]{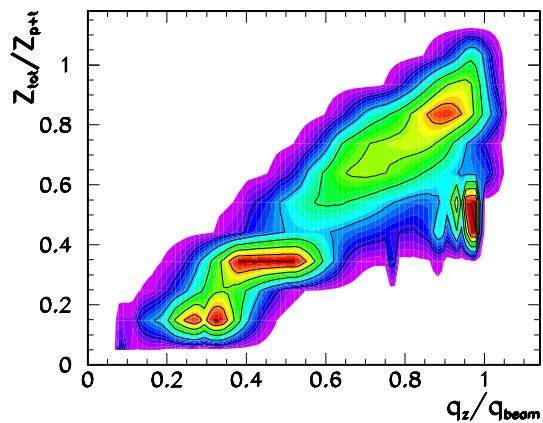}
\caption{Colour online: 
Contour plot of the total detected charge ($Z_{tot}$) normalized to the sum of
the projectile and target charge $Z_{p+t}$ as a function of the total measured
longitudinal momentum ($q_z$) normalized to the projectile momentum
($q_{beam}$). 
}
\label{ztot_qz}
\end{figure}

A first selection within the measured events is based on Figure~\ref{ztot_qz},
where we show the total detected charge as a function of the total longitudinal
momentum.
Requiring that at least 60\% of the total incoming parallel
momentum is collected, we obtain a total charge distribution centered
at $Z_{tot}=10$, corresponding to the 80\% of the total charge.
A yield peak around $Z_{tot}/ Z_{p+t}$ = 0.5 is evident in
the picture, corresponding to (quasi)$-$elastic events with only the
Carbon ejectile detected.
Since we would like to concentrate on specific decay channels, we
would keep a complete detection ($Z_{tot}=12$).
We have therefore checked that this stringent requirement does not
bias the characteristics of the events, comparing the
distribution of representative observables with a less stringent
selection $Z_{tot}\geq 10$.  Very similar distributions are
obtained with the two ``minimum bias'' selections which henceforth we
name ``quasi-complete'' ($Z_{tot}\geq 10$) and ``complete'' ($Z_{tot}=12$) (see
\S~\ref{Inclusive LCP energy spectra}).
Complete events are $\approx$ 20\% of quasi-complete ones.
\section{Data analysis and comparison to statistical model calculations}
With the minimum bias event selections discussed in
\S~\ref{Minimal bias event selection}, we compare
experimental data to  the predictions of our Monte Carlo
Hauser-Feshbach code HF$\ell$ (\S\ref{The statistical decay code})
for the evaporation of the CN $^{24}$Mg, at
$E^*/A_{CN}=2.6$ MeV, issued in case of complete fusion.
The angular momentum input distribution for the fused system in this
reaction can be assumed to be a triangular one, with a maximum
value $J_{0\ max}=12\ \hbar$, coming from the systematics~\cite{PACE}.
Because of parity conservation, only even values of $J_0$ extracted
from the triangular distribution are allowed as an input for the CN
angular momentum.
Finally, code predictions are filtered through a software replica of
the experimental set-up, taking into account the geometry, the
energy thresholds, the energy resolution and the solid angle for each detector.

The comparison of various experimental and simulated observables is
used to validate the parameterizations 
of statistical model ingredients implemented in the code.

\subsection{Experimental observables}
\label{Inclusive LCP energy spectra}

\begin{figure}
\centering
\includegraphics[angle=0,width=.7\columnwidth]{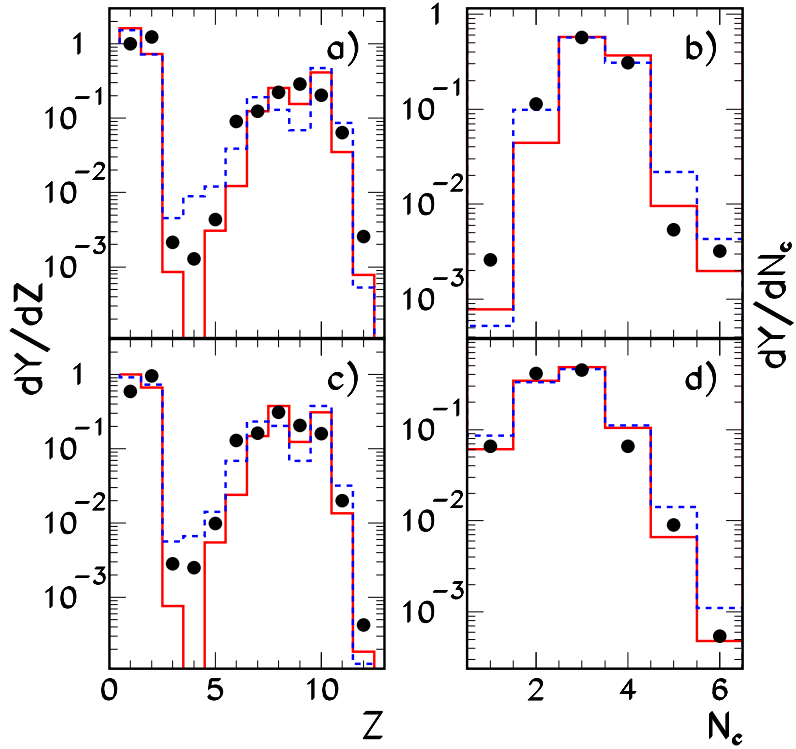}
\caption{Colour online: Measured (symbols) and calculated
  (histograms) charge and charged particle multiplicity distribution.
  Panels a), b) refer to complete ($Z_{tot}=12$) events, panels c), d)
  to quasi-complete events ($Z_{tot}\geq10$).
  The red solid line gives the result of the HF$\ell$ calculation, while the
  blue dashed line is obtained using the GEMINI++ model.
  All distributions are normalized to the total number of events.
 }
\label{zdist10-12}
\end{figure}
The inclusive charge and multiplicity distribution of events
completely and quasi-completely detected in charge are presented in
Figure~\ref{zdist10-12} in comparison with the filtered HF$\ell$
calculation. In this figure and in the following ones experimental 
data are always shown with statistical error bars, when visible.

The charge distribution is globally well-reproduced by the theoretical
calculation and its overall shape is typical of fusion-evaporation reactions.

However, a few discrepancies can be observed.
Notably, $Z=4$ fragments are absent in the Hauser-Feshbach
prediction while they are not  negligible in the experimental sample.
This could be interpreted as the presence of a break-up contribution
in the data which is not properly treated by the
sequential calculation. To confirm this statement, we show in the same figure
the result from a GEMINI++ calculation~\cite{gemini} subject to the same
filtering procedure.
This model, which has also been largely and successfully used by the nuclear
physics community since more than 20 years, includes the emission of
intermediate  mass fragments within the transition state formalism.
We can see that GEMINI++ predicts sizeable yields
of the lightest fragments, which in the
Hauser-Feshbach formalism have a negligible probability to be emitted
and are only  obtained as evaporation residues. In particular, the
transition state formalism succeeds in explaining the missing Be
cross section.

Concerning the multiplicity distribution, presented in the right
part of Figure~\ref{zdist10-12}, we can see that both the HF$\ell$ and the
GEMINI++ calculations reproduce the data satisfactorily. We can
however remark that GEMINI++ overpredicts events of high multiplicity. 
This means that the transition state formalism
is not entirely satisfactory in describing the production of light fragment,
which could also be due to a breakup mechanism. Such a mechanism is
not accounted for in the presented models.

Apart from the missing $Z=4$ channel, another discrepancy between the
HF$\ell$ calculation and the data concerns the $Z=6$ yield which is
underestimated by the model. This extra yield could in principle be
explained by the transition state model, as shown by the fact that data
are well reproduced in this channel by GEMINI++.
As an alternative explanation, the
Carbon excess with respect to HF$\ell$ predictions could be due to the
entrance channel of the reaction. Indeed, many other
experiments~\cite{nrv} where reactions 
with Carbon projectile and/or target were studied, showed an
extra-production of Carbon residues with respect to statistical models
expectations. At low bombarding energy, $C-C$ quasi-molecular
states~\cite{nrv} can be invoked. 
In our experiment, as it will be discussed in the following,
this anomaly is essentially associated with the specific $C-3\alpha$
channel.

Because of the great similarity between the HF$\ell$ and GEMINI++
calculations, and the fact that the HF$\ell$ code was optimized on light
systems (see \S\ref{The statistical decay code}),
in the following we exclusively use the HF$\ell$ code as a reference
statistical model calculation.

Due to the low statistics of the experiment for $Z=3, 4$ residues,
we will not study these residue channels any further.

\begin{figure}
\centering
\includegraphics[angle=0, width=.8\columnwidth]{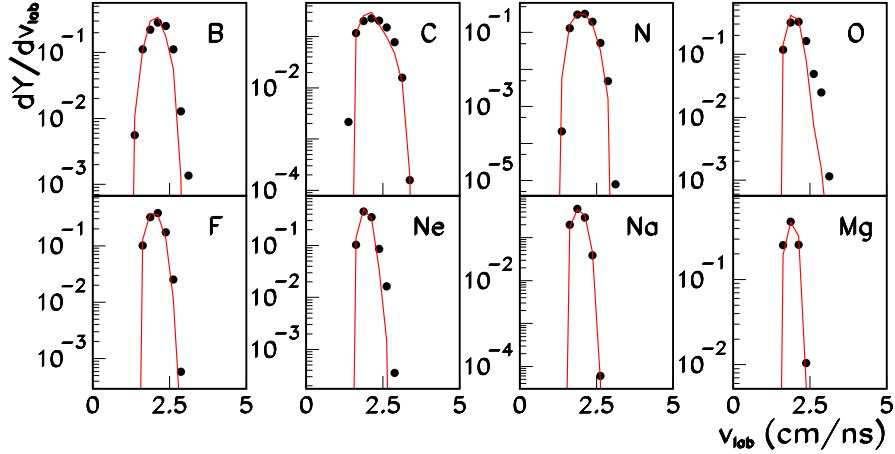}
\centering
\caption{(Colour online) Residue laboratory velocity for complete events
  ($Z_{tot}=12$) (symbols) in comparison with the HF$\ell$
  calculation (lines). All distributions are normalized to unitary area.
}
\label{vres_ztot12}
\end{figure}

The dominant fusion-evaporation character of the reaction is further
demonstrated in Figure~\ref{vres_ztot12}, which shows the velocity
distributions in the laboratory frame of the different fragments with
$Z\ge 5$. 
The good reproduction by the statistical model allows to
interpret these fragments mainly as evaporation residues left over by
the decay of $^{24}$Mg CN originated from complete fusion.

\begin{figure}
\centering
\includegraphics[angle=0, width=.7\columnwidth]{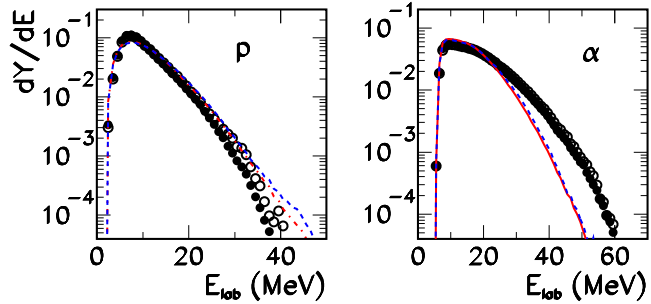}
\caption{
(Colour online) Proton (left part) and $\alpha$ (right part)
energy distributions  in the laboratory frame, detected
in quasi-complete $Z_{tot}\ge10$ events (full symbols) and in
complete $Z_{tot}=12$ events (open symbols).
Data (symbols) are compared to model calculations (lines).
Red solid lines: quasi-complete $Z_{tot}\ge10$ events.
Blue dashed lines: complete $Z_{tot}=12$ events.
All spectra are normalized to unitary area.
}
\label{dYdE_p_a_10_12}
\end{figure}

A complementary information is shown in Figure~\ref{dYdE_p_a_10_12},
which displays the laboratory energy spectra of protons and $\alpha$
particles detected in GARFIELD. 
Experimental data (dots) are compared to model calculations
(lines). 
From now on we will concentrate on events with a residue detected in
the RCo ($5^{\circ} \div 17^{\circ}$) and LCP in GARFIELD ($30^{\circ}
\div 150^{\circ}$) as in the previous experiment~\cite{prc}. 
This choice is essentially due to the different thresholds on LCP
identification in RCo and GARFIELD, as pointed out in \S\ref{The
experimental setup}. 
To facilitate the comparison of the spectral shapes, distributions are
always normalized to the same area. 
\begin{figure}
\centering
\includegraphics[angle=0, width=0.45\columnwidth]{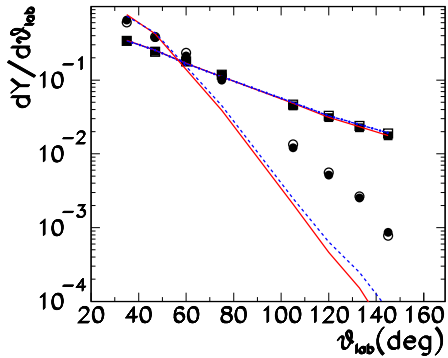}
\caption{(Colour online) Proton (squares) and $\alpha$ (circles)
 angular distributions in the laboratory frame, detected
 in complete $Z_{tot}=12$ (open symbols) and in quasi-complete
 $Z_{tot}\ge10$ (full symbols) events.
 Data (symbols) are compared to model calculations (lines).
 Blue dashed lines correspond to complete events, red solid lines to
 quasi-complete events. Experimental and calculated distributions are
 normalized to unitary area. 
 }
\label{angdist_p_a}
\end{figure}

We can see that the choice of the set of events (complete and
quasi-complete) does not deform the shape of 
the spectra. A satisfactory reproduction of the proton energy spectrum
is achieved, while a large discrepancy in the shape of the
distributions appears for $\alpha$ particles for both completeness
requirements.

Another information can be obtained from angular distributions of
protons and $\alpha$-particles (see Figure~\ref{angdist_p_a}). 
The proton distribution is in agreement with the model, while the
excess of $\alpha$ particles at backward laboratory angles could
suggest a preferential alpha emission from the quasi-target.
Alternatively, it could indicate an alpha transfer mechanism from an
excited $^{12}$C nucleus with strong alpha correlations.

As it is commonly known, the shape of LCP energy spectra is
determined by  the interplay of all physical ingredients entering in
the evaporation process,  notably including transmission coefficients,
angular momentum effects and level density~\cite{charity}.
Nevertheless, when comparing data to statistical model calculations, it is
possible to try to disentangle the effects of single
ingredients~\cite{charity}.
In particular, while transmission coefficients define the shape of evaporated
spectra in the Coulomb barrier region, the level density mostly affects
the slope of the exponential tail.
Concerning angular momentum, the inclusion of deformation has a stronger
influence on heavier fragment emission, as it is the case for
$\alpha$ particles, and, as a consequence, the tail of the energy
distribution for such fragments becomes steeper.

\begin{figure}
\centering
\includegraphics[angle=0, width=.7\columnwidth]{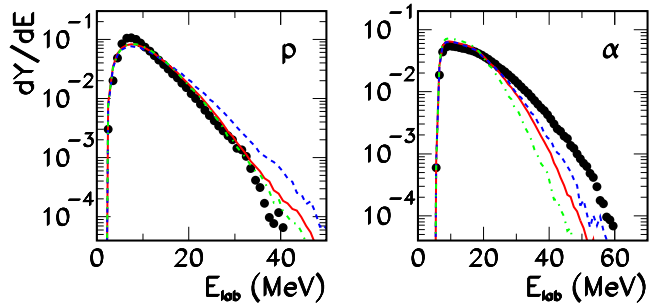}
\caption{
(Colour online) Proton (left part) and $\alpha$ (right part)
energy distributions  in the laboratory frame, detected
in  complete events.
Data (symbols) are compared to model calculations, with different
choices for the  assumed maximum angular momentum  $l_{max}$ and
limiting energy $E_l$.
Red solid lines: $l_{max}=12\hbar$, $E_l=3$ AMeV.
Blue dashed lines: $l_{max}=12\hbar$, $E_l=8$ AMeV.
Green dot-dashed lines: $l_{max}=6\hbar$, $E_l=3$ AMeV.
All spectra are normalized to unitary area.
}
\label{dYdE_p_a_theo}
\end{figure}

Thus the two theoretical uncertainties which could be responsible of
the observed deviations  are the estimated maximum angular momentum
leading to CN formation, and the level density
parameter. As we have discussed in \S~\ref{The statistical
 decay code}, the only unknown in the level density is the asymptotic
value of the $a$ parameter at very high excitation energy.
The effects of a very wide variation of these parameters, including an
unrealistically low value of $l_{max}$ and a very high
value of  $E_l$ are shown in Figure~\ref{dYdE_p_a_theo}.
This figure shows that no common choice on the LD parameters can be
done in order to reproduce at the same time proton and $\alpha$ energy
spectra.    For this reason we keep in the
rest of the analysis the fiducial values  $l_{max}=12\hbar$,
$E_l=3$ AMeV (red lines).

The comparison made so far on many inclusive observables suggests that
the dominant reaction mechanism is CN formation and  the discrepancy
found for $\alpha$ particles reflects an out-of-equilibrium  emission.

\begin{figure}
\centering
\includegraphics[angle=0, width=1\columnwidth]{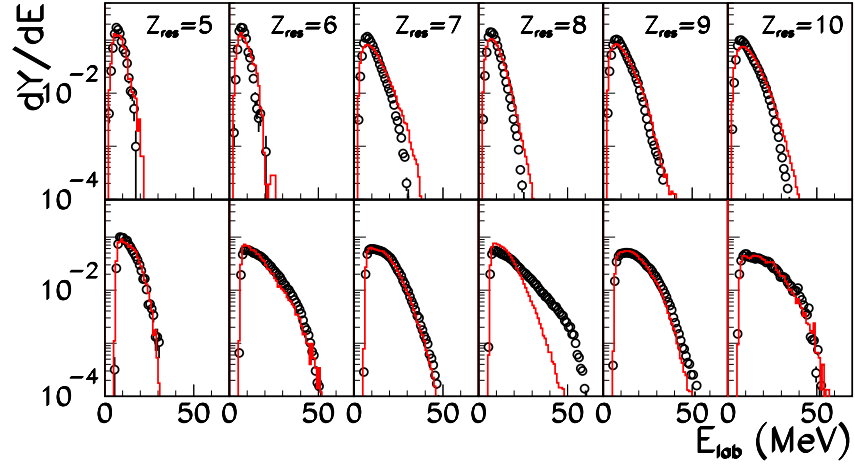}
\caption{(Colour online) Proton (upper part) and $\alpha$ (lower part)
  laboratory energy spectra in complete events detected in
coincidence with a residue of charge $Z_{res}$, indicated in each column.
Data (symbols) are compared to model calculations (lines).
All spectra are normalized to unitary area.}
\label{dYdE_residuo}
\end{figure}

A first confirmation of this hypothesis comes from the finding that
the largest source of disagreement between data and calculations is
for decay channels with $\alpha$ particles detected in
coincidence with an Oxygen fragment. 
This is shown in Figure~\ref{dYdE_residuo}, which presents energy
spectra of protons and $\alpha$ particles detected in coincidence with
a residue of a given atomic number. 
The discrepancy, larger at the most forward angles~\cite{baiocco}, is mainly
due to the 2 $\alpha$-channel, as we will discuss in
\S~{\ref{fully}}.

With the exception of the $\alpha-O$ coincidence, particle energy
spectra are very well reproduced by the statistical model.
\begin{figure}
\centering
\includegraphics[angle=0, width=.5\columnwidth]{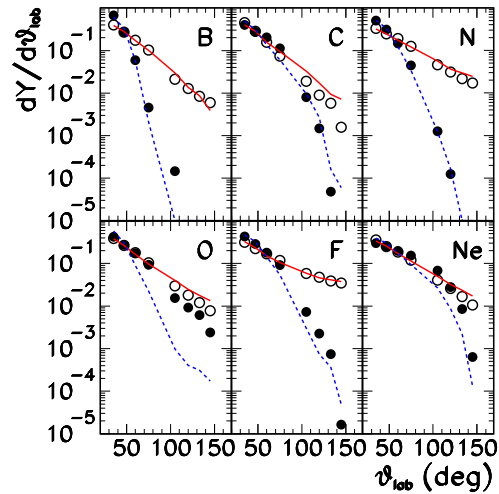}
\caption{ (Colour online)
Proton (open symbols) and $\alpha$ (full symbols)
angular distributions in the laboratory frame, detected in complete
events in coincidence with the indicated residues.
Data are compared to model calculations. Red solid lines correspond to
protons, blue dashed lines to $\alpha$'s.
Experimental and calculated distributions are normalized to unitary area.
 }
\label{angdist_p_a_res}
\end{figure}
This gives strong confidence to our level density model of
(\ref{rhou}) and (\ref{am}) with $E_l=3$ AMeV (corresponding to
$a\approx 3.5$ MeV$^{-1}$ for $E^*/A=3$ MeV) for the light $A\approx
20$ CN decay. 

A small difference of the experimental and calculated energy spectra 
is also observed for $\alpha$-particles in coincidence with a Carbon residue.
This does not seem to be related to the presence of peripheral events
with a Carbon quasi-projectiles, since the velocity distribution of
Carbon residues (shown in Figure~\ref{vres_ztot12}) displays a good
agreement with statistical calculations.

The angular distributions of protons and $\alpha$ particles in
coincidence with each residue are shown in Figure~\ref{angdist_p_a_res}.
The good agreement among data and model predictions as far as proton
distributions are concerned is confirmed. A large discrepancy is evident for
$\alpha$ particles at backward laboratory angles detected in
coincidence with an Oxygen fragment.

To understand the origin of the deviations from a statistical
behaviour, the branching ratios to $\alpha$ decay and $\alpha$
kinematics in the different channels involving $\alpha$ emission will
be studied in greater detail in Section \ref{fully}.

\subsection{Calorimetry and isotopic distributions}
\label{Calorimetry and isotopic distributions}

\begin{figure}
\centering
\includegraphics[angle=0, width=0.8\columnwidth]{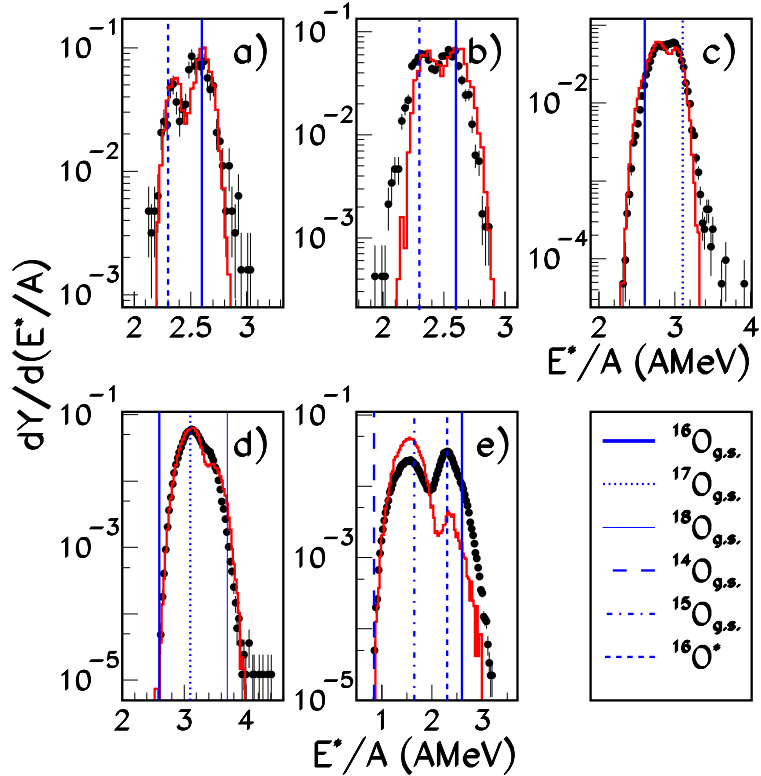}
\caption{(Colour online)  Calorimetric excitation energy distributions for the
different channels associated with the production of an Oxygen residue:
a) O+$\alpha$+2d; b) O+$\alpha$+p+t; c) O+$\alpha$+p+d; d) O+$\alpha$+2p;
e) O+2$\alpha$. 
Full symbols: experimental data; (red) lines: HF$\ell$
calculations. Blue vertical lines: expected values for the ground
(excited) states of different isotopes, as listed in the bottom right panel. 
Data and calculations are normalized to unitary area.
}
\label{fig_calo1}
\end{figure}

A deeper understanding of the reaction mechanism and a complementary
test of the statistical behaviour can be obtained
by studying the mass distribution of the different residues.
Unfortunately we do not have isotopic 
resolution for fragments with atomic number $Z\ge 5$.
However, if we consider in the analysis events completely detected in charge,
the residue mass can be evaluated
from the energy balance of the reaction, as we now explain.

Let us consider a well defined channel, characterized by a given residue charge
$Z_{res}$, light charged particle charge $Z_{lcp}=12-Z_{res}$ and mass
$A_{lcp}$. We define  $\bar{Q}=m_{lcp}c^2-m(^{24}Mg)c^2 $ the {\it partial}
$Q$-value associated with that channel, where
$m_{lcp}$ is the total mass of the channel particles and $m(^{24}Mg)$ is the
mass of the composite nucleus.
The unknown residue mass number $A_{res}^k$ and unknown neutron number $N_n^k$ in
each event $k$ belonging to the considered channel are defined as a function
of an integer isotopic variable $x$ as
$A_{res}(x)=2Z_{res}+x$ and $N_n(x)=24-A_{lcp}-A_{res}(x)=24-A_{lcp}-2Z_{res}-x$.
The residue mass and total neutron energy  are thus defined as a
function of $x$: 
$m_{res}(x)=m(A_{res}(x),Z_{res})$, $E_n(x)=(\langle
e_n\rangle+m_nc^2)N_n(x)$, where $\langle e_n\rangle$ is the estimate of
the average neutron kinetic energy from the average measured proton
one, with the subtraction of an average $2.9$ MeV Coulomb barrier. 

 The excitation energy of the event $k$  reads:
\begin{equation}
E^*_{theo}=\bar{Q}+m_{res}(x)+E_{kin}^k+E_n(x)+E_\gamma^k
\label{calo1}
\end{equation}
where $E_{kin}^k$ is the total measured kinetic energy, $E^*_{theo}=62.4$ MeV
the total available energy, and $E_\gamma^k$ the
unmeasured $\gamma$ energy, in the centre of mass system.
The excitation energy which would be associated to this event assuming
that the residue has $A_{res}=2Z_{res}$ and is produced in its ground state is:
\begin{equation}
E_{cal} (k)=\bar{Q}+m_{res}(x=0)+E_{kin}^k+E_n(x=0)
\label{calo2}
\end{equation}
The example of Oxygen is reported in Figure~\ref{fig_calo1}.
The calorimetric excitation energy distribution
$E_{cal}(k)$ divided by the total mass of the system is displayed
for the different measured
channels associated to the production of $Z=8$ fragments in complete
events, together with the filtered model calculations. 
In all cases we can observe a
wide distribution corresponding to different, often unresolved 
states of different isotopes.
The qualitative agreement with the model calculations confirms once
again that the selected events largely correspond to complete
fusion. 

In the hypothesis that the kinetic energies of LCP and neutrons depend
on average on the channel, but not on the average value of
the residue mass (through $\langle x\rangle$), (\ref{calo1})
and~(\ref{calo2}) can be averaged over the events of the channel giving:
\begin{eqnarray}
E^*_{theo}&=&\bar{Q}+m_{res}(x)+\langle E_{kin}\rangle+E_n(x)+\langle E_\gamma\rangle \\
\langle E_{cal}\rangle&=&\bar{Q}+m_{res}(x=0)+\langle E_{kin}\rangle+E_n(x=0)
\end{eqnarray}

Eqs. (10) and (11) allow for deducing the unmeasured neutron excess, and
therefore the residue mass, from the average measured calorimetric
energy. Indeed, subtracting the two equations we get:
\begin{equation}
\fl \langle E_{cal}\rangle(x,I) = E^*_{theo}-(m_{res}(x)-m_{res}(x=0))-(E_n(x)-E_n(x=0))-E^*_I
\label{calo3}
\end{equation}

This equation gives the calorimetric energy which is expected in average for
a residue  of mass number $A_{res}=2Z_{res}+x$ produced in its excited
state $I$, if we have assumed via (\ref{calo2}) that its mass number
is $2Z_{res}$, as shown for various cases by the blue vertical lines
in Figure~\ref{fig_calo1}.

Our energy resolution is not sufficient to determine the detailed
spectroscopy of each residue, but the comparison of the measured
calorimetric energy in each event given by (\ref{calo2}) with the expected
value from~(\ref{calo3}) allows for a reasonably good isotopic
identification.

\begin{figure}
\centering
\includegraphics[angle=0, width=1\columnwidth]{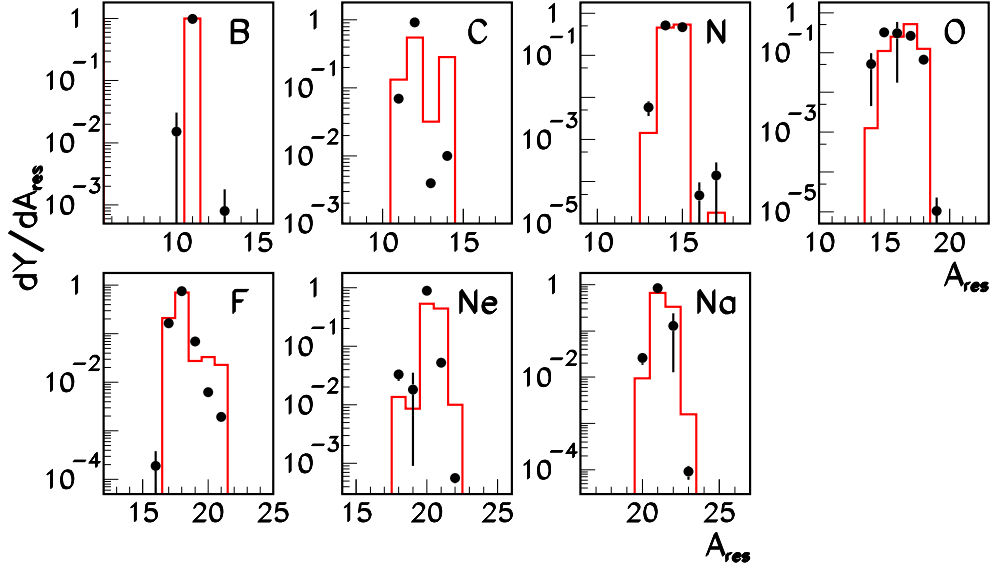}
\caption{(Colour online)  Experimental (symbols)
isotopic distribution of residues obtained for complete
events using~(\ref{minim}) (see text), compared to theoretical
predictions (histograms). 
Spectra are normalized to unitary area.} 
\label{fig_calo3}
\end{figure}

To attribute a
definite isotope to each residue, we have minimized
in each event $k$ the distance in energy between the calorimetric
result and the theoretical value associated to the resolved states  of
the associated channel
\begin{equation}
|E_{cal}(k)-\langle E_{cal}\rangle(x,I)|=min
\label{minim}
\end{equation}
We have repeated the same procedure for all the residues.
The resulting isotopic distributions are presented in Figure~\ref{fig_calo3},
again compared to the model calculations.
Errors on experimental results have been obtained combining the
statistical error with the one coming from the reconstruction
procedure. This has been estimated by comparing, within the model,
the values obtained by the reconstruction procedure with the original
predictions. 
The global agreement is good, particularly for odd charge residues.
Both the average and the width of the distributions are reproduced by the
model. The distributions are generally bell-shaped and structureless, with
the exception of Carbon, which shows an important depletion for
$^{13}$C similarly to the model calculation.

The case of Oxygen is particularly interesting. The experimental and
theoretical widths are comparable, but while the experimental distribution
has a negative skewness and it is centered on the neutron poor $^{15}$O,
the opposite is seen in the calculation which favours neutron rich isotopes
and presents a positive skewness~\cite{prc}.
As we can see in Figure~\ref{fig_calo1}, this is largely due to the
specific $O+2\alpha$ channel.
Indeed this channel is the only one which leads to a non-negligible production
of $^{15}$O.

The information from the isotopic distribution and the energy spectra
coherently points towards an increased probability for the
$O+2\alpha$ channel with respect to the  statistical model.
We therefore turn to see if the experimental sample
contains, together with a dominant contribution of standard compound
reactions, other reaction mechanisms  which could selectively populate
a few specific channels, possibly associated with $\alpha$ emission.

\subsection{Multiple $\alpha$ channels}
\label{fully}

In Table~\ref{TAB1} we report for each residue the most populated
channel in the experimental sample, as well as the associated
branching ratio. The results are compared to the prediction of the
statistical model for the same channel, filtered through the
characteristics of the experimental apparatus. 
We can see that the branching ratio of the dominant decay channels is
reasonably well reproduced by the statistical model for odd-Z
residues, while discrepancies can be seen for even-Z ones. 

\begin{table}[h]
\begin{center}
\begin{tabular}{|c|c|c|c|}
\hline
 $Z_{res}$ &   channel  &$BR_{\ HF\ell}$ & $BR_{\ EXP}$ \\
\hline
5 &   $^{11-xn}B$+$xn$+$p$+3$\alpha$    &100\%     &99\%    \\
6 &    $^{12-xn}C$+$xn$+3$\alpha$   &66\%     &98\%    \\
7 &   $^{15-xn}N$+$xn$+$p$+2$\alpha$   &94\%     &91\%    \\
8 &    $^{16-xn}O$+$xn$+2$\alpha$    &11\%      &63\%    \\
9 &   $^{19-xn}F$+$xn$+$p$+$\alpha$     &87\%      &92\%     \\
10 &   $^{22-xn}Ne$+$xn$+2$p$ &84\% &55\%  \\
\hline
\end{tabular}
\vspace{.5cm}
\caption{For each measured residue, the table gives the most probable
experimental channel and its branching ratio together with the value predicted
by the HF$\ell$ calculations. Errors on the experimental
values (about 5\%) take into account both the statistical error and the
possible $^3$He-$\alpha$ contamination.
}
\label{TAB1}
\end{center}
\end{table}

\begin{figure}
\centering
\includegraphics[angle=0, width=0.7\columnwidth]{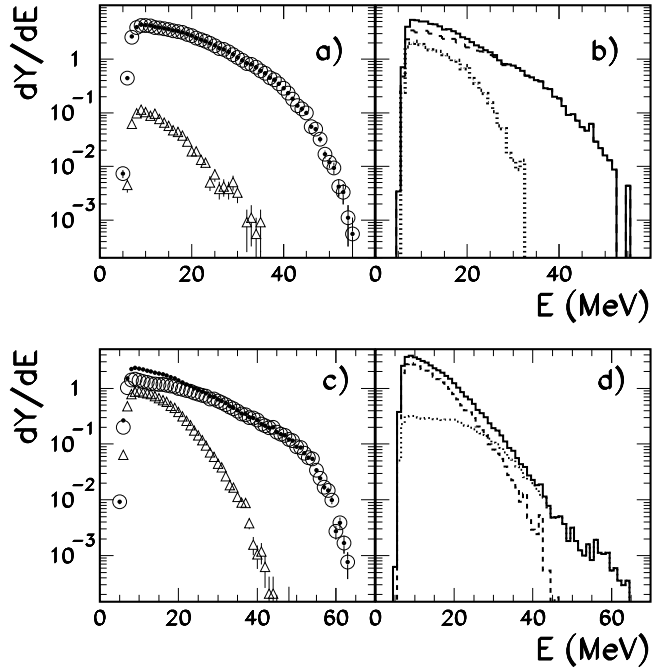}
\caption{Experimental energy spectra (left) compared to HF$\ell$
calculations (right). 
Upper part: Carbon residue: full symbols and full line represent the
inclusive distribution of all decay channels, open triangles and
dashed line correspond to channels involving two $\alpha$'s and two
hydrogens, open circles and dotted line correspond to the three
$\alpha$'s channel. 
Lower part: Oxygen residue: Full symbols and full line represent the
inclusive distribution of all decay channels, 
open triangles and dashed line correspond to $O+\alpha+2H$ channel
and open circles and dotted line correspond to $O+2\alpha$ channel.
The spectra are normalized to the number of events of each residue.}
\label{C+Oa_br}
\end{figure}
\begin{figure}
\centering
\includegraphics[angle=0, width=0.8\columnwidth]{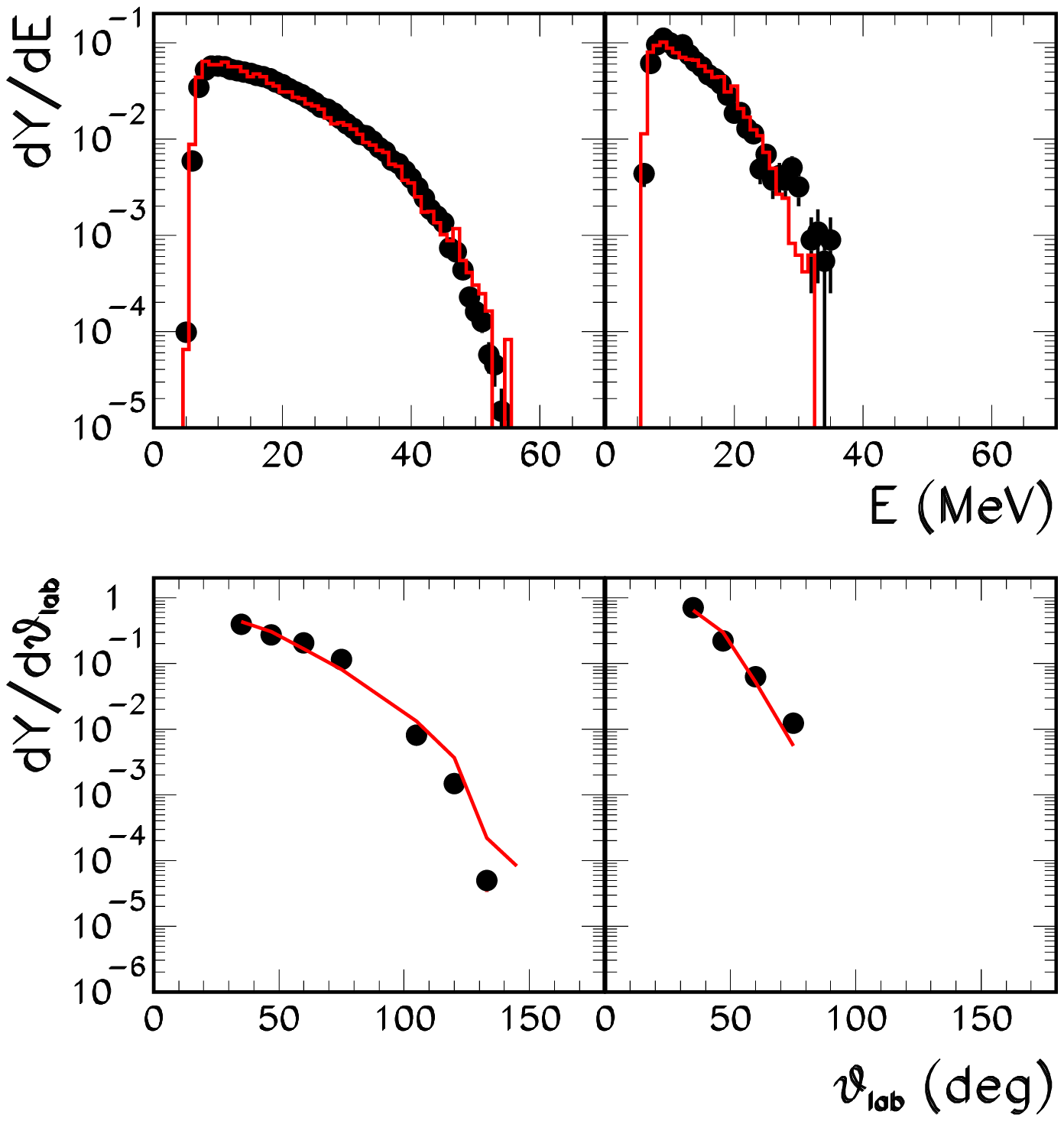}
\caption{(Colour online) 
Energy spectra (upper panels) and angular distributions (lower panels)
of $\alpha$-particles detected in coincidence with a Carbon residue.
Data (symbols) are compared to HF$\ell$ calculations (lines) for the two
channels $C+3\alpha$  (left) and $C+2\alpha+2H$ (right). All the spectra are
normalized to unitary area in order to compare the shapes
independently of the different branching ratios.}
\label{Cac_br}
\end{figure}
\begin{figure}
\centering
\includegraphics[angle=0, width=0.8\columnwidth]{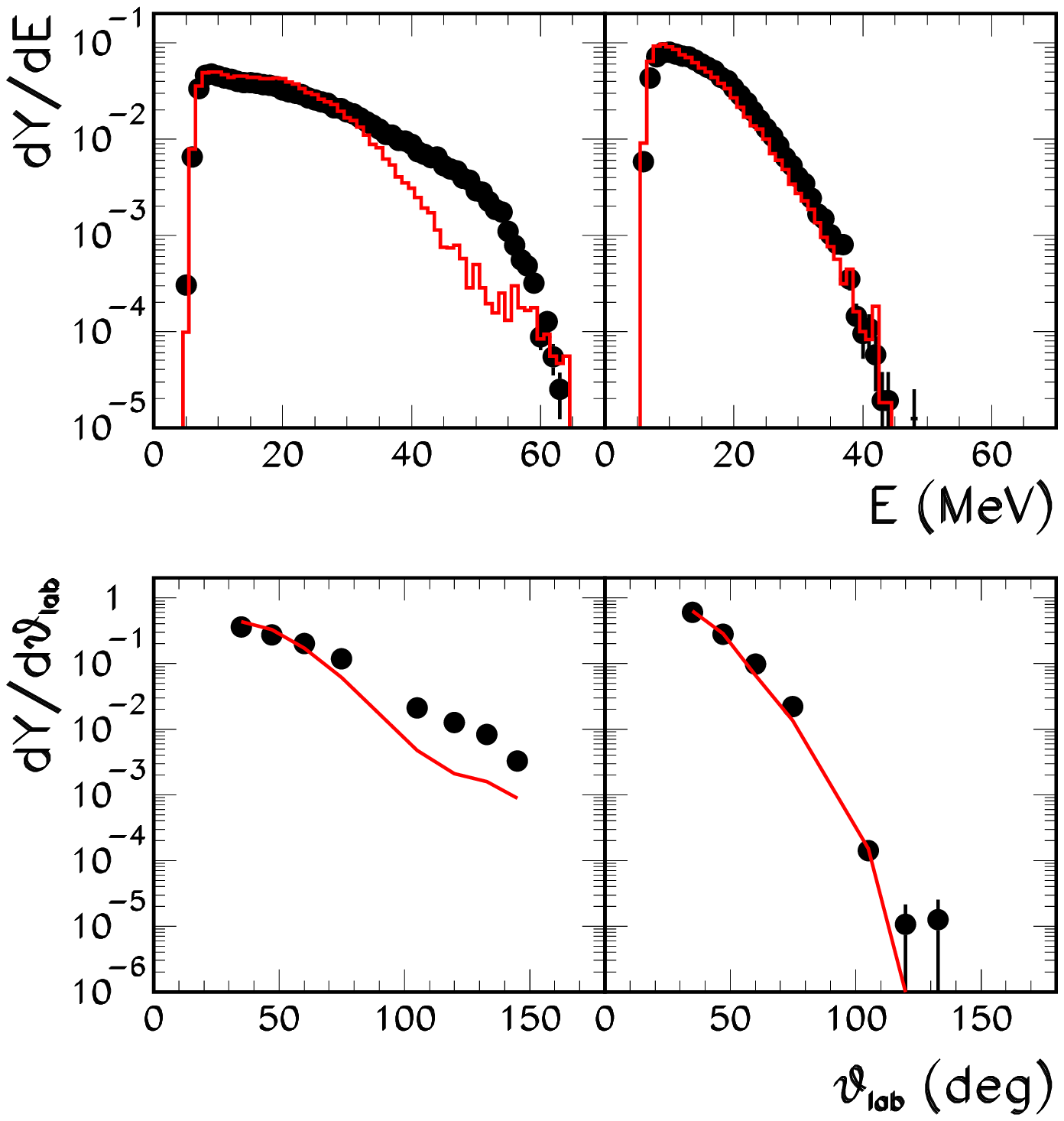}
\caption{(Colour online) 
Energy spectra (upper panels) and angular distributions (lower panels)
of $\alpha$-particles detected in coincidence with an Oxygen residue.
Data (symbols) are compared to HF$\ell$ calculations (lines) for the two
channels $O+2\alpha$  (left) and $O+\alpha+2H$ (right). All the spectra are
normalized to unitary area in order to compare the shapes
independently of the different branching ratios.}
\label{Oac_br}
\end{figure}

For Oxygen, the predicted most probable channel is
$^{A}$O+$\alpha$+2$H$ (here 2H stands for two Z=1 products)
with a branching ratio $BR_{\ HF\ell}=88\%$, while this
channel is experimentally populated  with  $BR_{\ EXP}=37\%$.
For Carbon, $^{12}$C+3$\alpha$ is the most probable theoretical
channel consistent with the data, but an important contribution of the
channel $^{12}$C+2$\alpha$+$2H$ is also predicted ($BR_{\ HF\ell}=32\%$),
while this contribution is negligible in the experimental sample.
Also for Neon a disagreement is present,
but the theoretical calculation well reproduces the shape of the
$\alpha$ spectrum, as shown in Figure~\ref{dYdE_residuo}. This is not
the case for Oxygen and, to a lesser extent, Carbon. For these
residues the discrepancy in the branching ratios affects the shape of
the $\alpha$ particle spectra. 
This is shown in Figure~\ref{C+Oa_br}, where, for the Carbon
case (upper panels), the measured inclusive $\alpha$ spectrum is
dominated by the multiple $\alpha$ channel, while in the statistical
model the channel containing only two $\alpha$ particles (and hence
two hydrogen isotopes) is very important 
for low $\alpha$ energies, thus modifying the slope of the inclusive 
spectrum with respect to the data.

A similar analysis for Oxygen is presented in Figure~\ref{C+Oa_br}
(lower panels). 
The same considerations as for Carbon apply in this case. Again,
the extra yield associated with multiple $\alpha$ events with respect to
the statistical model leads to a broader spectrum extending towards
higher $\alpha$ energies.

If we now compare experimental data with model predictions in specific
channels, we obtain, for the Carbon case (see Figure~\ref{Cac_br},
upper panels), that the shape of the spectra of the different channels 
are very well reproduced by the statistical model calculations. The
same holds true for the angular distributions, well reproduced by
calculations.
This shows that the kinematics of the decay is well described by a
sequential evaporation mechanism. 
However these shapes depend on the channel, multiple $\alpha$'s  leading
to spectra which are less steep and extend further in energy, with respect
to channels where hydrogens are also present.
Because of that, the disagreement in branching ratios between model
and data shown in Table I affects the global shape of the
$\alpha$ spectrum, where the different channels are summed up.
 
Taking now into account the Oxygen residue, in Figure~\ref{Oac_br}
we show the comparison of the energy spectra (upper panels) and
angular distributions.  At variance with the Carbon case, the shape of
the $\alpha$ spectrum and the angular distribution in the $O+2\alpha$
channel are not well reproduced  by the statistical model. This means
that the kinematics in this channel is not compatible  with CN decay,
and suggests a contamination from direct reactions.

The anomalously high  probability of multiple $\alpha$ emission in
coincidence with Oxygen and Carbon residues, with respect to
the expectation from a statistical behaviour,
can explain the deviations observed in the inclusive $\alpha$
observables (see Figures~\ref{dYdE_p_a_10_12} and \ref{angdist_p_a}).
This suggests that non-statistical processes are at play in the
experimental sample concerning the two specific multiple $\alpha$
channels that show anomalously high  branching ratios.

Alpha production is known to be an important outcome of direct
$^{12}$C+$^{12}$C reactions~\cite{inelastic,transfer}. In these studies,
though at lower bombarding energy than the present experiment, the
$\alpha$ dominance has been associated with
quasi-molecular two-Carbon excited states with a pronounced $\alpha$
structure.

In order to see if similar effects still persist at higher bombarding energies,
in the second paper of the series we will focus on a detailed analysis
of the multiple $\alpha$ channels.

\section{Conclusions}
In this work we have presented results for the $^{12}$C($^{12}$C,X) reaction at
95 MeV beam energy, measured at LNL-INFN with the GARFIELD+RCo experimental
set-up.

Starting from a minimal selection of the fusion-evaporation mechanism, based
on the coincidence between LCP's emitted over a wide polar angle range
(GARFIELD) and a fragment detected at laboratory forward angles (RCo),
reinforced by completeness conditions on the total detected charge and
longitudinal momentum, we have compared experimental data to statistical
model calculations for the decay of the $^{24}$Mg$^*$ CN issued in case of
complete fusion.

The selected sample is compatible with the expected behaviour of
a complete-fusion-evaporation reaction, with the exception of two specific
channels significantly more populated than predicted
by the HF$\ell$ calculations.
These channels  correspond to the emission of two or three $\alpha$
particles in coincidence with an Oxygen or Carbon residue, respectively.
The $\alpha$ spectra and angular distributions in the $(O+2\alpha)$ channel
are not compatible with statistical model calculation. This suggests a
contamination from direct reactions or $\alpha$-structure correlations
in the $^{24}$Mg compound~\cite{inpc13}.

This is not the case for the $(C+3\alpha)$ channel, and the
anomalously high branching ratio of this channel can be tentatively
attributed to a possible persistence at high excitation energy of $\alpha$
structure correlations in the $^{12}$C+$^{12}$C molecular state and/or
in the $^{24}$Mg compound. 
The kinematic characteristics of these
non-statistical decays are  further studied in the continuation of
this work~\cite{next}.

The results of the analysis show that our data can be used to constrain
the ingredients of the statistical model in the $A\leq 24$,
$E^{*} \leq 2.6$ AMeV mass-excitation-energy region of interest.

In particular, this analysis supports a 
model showing a very steep increase of the level density with
excitation energy. 
The value of the level density parameter around $3$ AMeV 
excitation energy extracted from this study
is consistent with early findings from fragmentation experiments.

\ack
The authors thank the crew of the XTU TANDEM acceleration system at LNL.
\\
This work was partially supported by the European Funds for Large
Scale Facilities - Seventh Framework Program - ENSAR 262010 and by
grants of Italian Ministry of Education, University and Research under
contract PRIN 2010-2011.

\end{document}